\documentclass[pra,aps]{revtex4}
\usepackage{amssymb}
\usepackage{amsfonts}
\usepackage{amsmath}
\usepackage{graphicx}

\DeclareMathOperator{\Tr}{Tr}
\DeclareMathOperator{\arctanh}{arctanh}
\newcommand{\beq}{\begin{equation}}
\newcommand{\eeq}{\end{equation}}

\newcommand{\bea}{\begin{eqnarray}}
\newcommand{\eea}{\end{eqnarray}}

\begin{document}

\title{Error estimation in the direct state tomography}
\author{I. Sainz }
\affiliation{Departamento de F\'{\i}sica, Universidad de Guadalajara, Revoluci\'{o}n 1500, 44420
Guadalajara, Jal., M\'exico.}
\author{A. B. Klimov }
\affiliation{Departamento de F\'{\i}sica, Universidad de Guadalajara, Revoluci\'{o}n 1500, 44420
Guadalajara, Jal., M\'exico.}
\date{\today }

\begin{abstract}
We show that reformulating the Direct State Tomography (DST)
     protocol in terms of projections into a set of non-orthogonal
bases one can perform an accuracy analysis of DST in a similar way as
in the standard projection-based reconstruction schemes. i.e. in terms
of the Hilbert-Schmidt distance between estimated and true states.
This allows us to determine the estimation error for any measurement
strength, including  the weak measurement case, and to obtain an explicit analytic  form for the average minimum square errors.
\end{abstract}

\maketitle

\section{Introduction}

An appealing physical idea of the weak measurement tomography \cite%
{johansen,lundeen1,lundeen2,elefante,salvail} offers a possibility of
reconstruction of the wave function in a single experimental setup that
involves a specific system-pointer coupling; the so-called Direct State
Tomography (DST). Basically, the scheme consists in  successive
measurements of two complementary observables of the system, where only the
first one is weakly coupled to the measurement apparatus \cite%
{lundeen1,lundeen2,elefante,salvail}. Recently, this procedure was
generalized to arbitrary coupling strengths \cite{vallone}, and it was
argued that the strong measurements expectably outperform the weak ones both
in precision and accuracy.

Since in the framework of weak measurements the efficiency is traded for accuracy, the error estimation analysis  becomes vital. Typically, the experimental performance of DST in the case of weak
measurements, is compared either with results of strong (projective)
tomography \cite{lundeen1,salvail}. An alternative method of estimation of the fidelity of a
reconstructed state was introduced in \cite{das}. However,  non of the above-mentioned approaches  analyzed the global intrinsic error estimation \cite%
{helstrom,englert,checos}.

In this letter we show that conveniently reformulating the approach \cite%
{vallone} as a Mutually Unbiased Bases (MUB)-like reconstruction scheme in
non-orthogonal bases \cite{pra,jpm} one can carry out the mean square error
(MSE) analysis of DST , including the weak measurement limit, in the
framework of measurement statistics \cite{englert,checos}. In particular, we
exemplify on the single qubit case that non-orthogonal bases appear as effective
projective states, in such a way that a weak coupling corresponds to
projection into near-parallel bases.
This allows us to reformulate the accuracy analysis in terms of measured probabilities. And thus, estimate the intrinsic statistical errors finding the minimum MSE using the Cr\'amer-Rao lower bound.

\section{Direct state tomography and non-orthogonal bases}

Following the general idea of DST we consider an unknown state $\rho _{s}$
of the system (one qubit) interacting with a pointer (another qubit)
initially prepared in the eigenstate state $\left\vert 0\right\rangle
\left\langle 0\right\vert _{p}$ of the Pauli operator $\sigma _{zp}$,
according to
\begin{equation*}
U(\theta )=e^{-i\theta \sigma _{xs}\otimes \sigma _{xp}},
\end{equation*}%
where $\theta \in \lbrack 0,\pi /4]$ is the measurement strength. After the
interaction the system is postselected in the state $\left\vert
0\right\rangle _{s}$, and the pointer is measured in the bases $\{\left\vert
\mathbf{e}^{t}\right\rangle ,t=0,1,2\}$:
\begin{eqnarray}
\left\vert e_{0}^{0}\right\rangle =\left\vert 0\right\rangle , &&\left\vert
e_{1}^{0}\right\rangle =\left\vert 1\right\rangle ,  \notag \\
\left\vert e_{0}^{1}\right\rangle =\frac{1}{\sqrt{2}}(\left\vert
0\right\rangle +\left\vert 1\right\rangle ), &&\left\vert
e_{1}^{1}\right\rangle =\frac{1}{\sqrt{2}}(\left\vert 0\right\rangle
-\left\vert 1\right\rangle ),  \notag \\
\left\vert e_{0}^{2}\right\rangle =\frac{1}{\sqrt{2}}(\left\vert
0\right\rangle -i\left\vert 1\right\rangle ), &&\left\vert
e_{1}^{2}\right\rangle =\frac{1}{\sqrt{2}}(\left\vert 0\right\rangle
+i\left\vert 1\right\rangle ),  \notag
\end{eqnarray}%
and the following probabilities are retrieved
\begin{equation}
N_{kt}^{2}\left\langle e_{k}^{t}\right\vert _{p}\left\langle 0\right\vert
_{s}U(\theta )\rho _{0}U^{\dag }(\theta )\left\vert 0\right\rangle
_{s}\left\vert e_{k}^{t}\right\rangle _{p}=p_{kt},  \label{pik}
\end{equation}%
where $\rho _{0}=\rho _{s}\otimes \left\vert 0\right\rangle \left\langle
0\right\vert _{p}$ and $N_{kt}$ are the normalization constants. In the
framework of DST \cite{lundeen1}, \cite{vallone} the wave function is
reconstructed in the basis $\{\left\vert e_{k}^{1}\right\rangle \}$ of
eigenstates of $\sigma _{xs}$ as a linear combination of the probabilities $%
p_{kt}$. On the other hand, the probabilities (\ref{pik}) can be considered
as projections of the initial state $\rho _{0}$ into the set

\begin{equation}
\left\vert \psi _{k}^{t}\right\rangle _{s}=N_{kt}\left\langle 0\right\vert
_{p}U^{\dag }(\theta )\left\vert 0\right\rangle _{s}\left\vert
e_{k}^{t}\right\rangle ,  \label{psiik}
\end{equation}%
so that
\begin{equation}
p_{kt}=\left\langle \psi _{k}^{t}\right\vert \rho _{s}\left\vert \psi
_{k}^{t}\right\rangle _{s}.  \label{pkt}
\end{equation}%
It is worth noticing that similar effective projection states naturally appear
in experiments \cite{mdr,shikano}.

Explicitly, the effective projection states (\ref{psiik}) have the form
\begin{eqnarray}
\left\vert \psi _{0}^{0}\right\rangle _{s} &=&\left\vert 0\right\rangle
_{s},\quad \left\vert \psi _{1}^{0}\right\rangle _{s}=\left\vert
1\right\rangle _{s}  \notag \\
\left\vert \psi _{k}^{1}\right\rangle _{s} &=&\cos \theta \left\vert
0\right\rangle _{s}+(-1)^{k}i\sin \theta \left\vert 1\right\rangle _{s}
\notag \\
\left\vert \psi _{k}^{2}\right\rangle _{s} &=&\cos \theta \left\vert
0\right\rangle _{s}+(-1)^{k}\sin \theta \left\vert 1\right\rangle _{s},
\label{bases}
\end{eqnarray}%
where $k=0,1$, and satisfy the condition
\begin{equation*}
|\langle \psi _{0}^{t}|\psi _{1}^{t}\rangle |=\cos 2\theta ,\;t=1,2,
\end{equation*}%
defining the so-called equidistant bases \cite{petal}.

Introducing $\lambda =\cos 2\theta $ we rewrite elements $\left\vert \psi
_{k}^{t}\right\rangle _{s}$ as follows
\begin{equation}
\left\vert \psi _{k}^{t}(\lambda )\right\rangle _{s}=\sqrt{\frac{1+\lambda }{%
2}}\left\vert 0\right\rangle _{s}+(-1)^{k}(-i)^{t+2}\sqrt{\frac{1-\lambda }{2%
}}\left\vert 1\right\rangle _{s},  \label{psil}
\end{equation}%
for $t=1,2$ so that $\left\vert \psi _{k}^{t}(\lambda =0)\right\rangle
_{s}=\left\vert e_{k}^{t}\right\rangle $ and the limit $\lambda \rightarrow
1 $ of almost \textquotedblleft parallel\textquotedblright\ states (close to
$\left\vert 0\right\rangle $) corresponds to the weak measurement case, $%
\theta \rightarrow 0$.

An important feature of the probabilities $p_{kt}$, $t=1,2$ is the relation
\cite{pra}
\begin{equation}
p_{0t}+p_{1t}=1-\lambda +2\lambda p_{00}=S,  \label{s}
\end{equation}%
which reflects a statistical dependence on the measurements in the
non-orthogonal and computational bases.

The bases (\ref{bases}) form an informationally complete set for $|\lambda
|<1$ \cite{petal}, and the density matrix of the system can be reconstructed
in terms of the probabilities (\ref{pkt}) according to \cite{pra}
\begin{eqnarray}
\rho _{s} &=&\frac{1}{1-\lambda ^{2}}\sum_{t=1}^{2}\sum_{k=0}^{1}p_{kt}\left%
\vert \phi _{k}^{t}\right\rangle \left\langle \phi _{k}^{t}\right\vert
\label{reconstruccion} \\
&&+\frac{1-\lambda }{1+\lambda }(p_{00}-1)\left\vert \psi
_{0}^{0}\right\rangle \left\langle \psi _{0}^{0}\right\vert -\frac{1+\lambda
}{1-\lambda }p_{00}\left\vert \psi _{1}^{0}\right\rangle \left\langle \psi
_{1}^{0}\right\vert ,  \notag
\end{eqnarray}%
where $\left\{ \left\vert \phi _{k}^{t}(\lambda )\right\rangle =\left\vert
\psi _{k}^{t}(-\lambda )\right\rangle \right\} $, is the corresponding $t$%
-th biorthogonal basis, $\langle \phi _{k}^{t}|\psi _{l}^{t}\rangle =\sqrt{%
1-\lambda ^{2}}\delta _{kl}$. In the limit $\lambda =0$ ($\theta =\pi /4$)
the expression above is converted to the standard (orthogonal) MUB
tomographic expression \cite{optimal} allowing the maximum
information gain \cite{wootters}.

\section{Error estimation in direct state tomography}

Each projector in the set $\{\left\vert \psi _{k}^{t}\right\rangle
\left\langle \psi _{k}^{t}\right\vert ,t=0,1,2;k=0,1\}$ can be considered as
a single output channel of an effective measuring apparatus. An estimation
procedure consists in a repetitive measurement on each of $N$ identical
copies of the system, i.e. the pointer is postselected in every basis the
same number of times, obtaining frequencies $\nu _{kt}=n_{kt}/N$, where $%
n_{kt}$ is the number of projections into $\left\vert \psi
_{k}^{t}\right\rangle $. The corresponding statistics of outcomes is
binomial-like \cite{preparacion}
\begin{equation}
P_{t}(\mathbf{n_{t}}|\mathbf{p_{t}})=\frac{1}{S^{N}}\frac{N!}{n_{0t}!n_{1t}!}%
p_{0t}^{n_{0t}}p_{1t}^{n_{1t}},  \label{binomial}
\end{equation}%
where $\mathbf{n_{t}}=(n_{0t},n_{1t})^{T}$, $\mathbf{p_{t}}%
=(p_{0t}p_{1t})^{T}$, $N=n_{0t}+n_{1t}$ and the condition (\ref{s}) is
satisfied. For the computational basis, where $S=1$, the statistics is
obviously binomial .

The expectation values corresponding to the probability distribution (\ref%
{binomial}) are of the form
\begin{equation}
\langle n_{kt}\rangle =\frac{Np_{kt}}{S},~~\langle n_{kt}^{2}\rangle =\frac{%
Np_{kt}}{S^{2}}\left[ (N-1)p_{kt}+S\right] .  \label{expectation}
\end{equation}

Following general ideas \cite{englert,checos}, we compute the estimation
error as the average squared of the Hilbert-Schmidt distance between the
true $\rho _{s}$ and estimated $\hat{\rho}_{s}$ system states,
\begin{equation}
\langle \mathcal{E}^{2}\rangle =\langle Tr[(\rho -\hat{\rho})^{2}]\rangle .
\label{E}
\end{equation}
It depends on $\lambda $ and the inner product between all the projectors
appearing in (\ref{reconstruccion}).

Taking into account (\ref{s}) we obtain for the difference between true ($%
p_{kt}$) and estimated ($\hat{p}_{kt}$) probabilities, $\Delta p_{kt}=p_{kt}-%
\hat{p}_{kt}$, the following relations
\begin{eqnarray*}
\Delta p_{10} &=&-\Delta p_{00}, \\
\Delta p_{1t} &=&2\lambda \Delta p_{00}-\Delta p_{0t},\;t=1,2.
\end{eqnarray*}%
Substituting the above relations into (\ref{E}) and averaging for many
repetitions, we obtain for the average quadratic error
\begin{equation}
\left\langle \mathcal{E}^{2}\right\rangle =\sum_{t,r=0}^{2}q_{tr}\langle
\Delta p_{0t}\Delta p_{0r}\rangle ,  \label{cost}
\end{equation}%
where the\ explicit form of the coefficients $q_{tr}$ (in matrix form) is
given in Appendix. Employing the Cram\'{e}r-Rao lower bound we minimize the
possible mean square error (MSE) per trail \cite{helstrom}
\begin{equation}
\left\langle \mathcal{E}^{2}\right\rangle \geq \Tr(QF^{-1}),  \label{mse}
\end{equation}%
where $Q=[q_{tr}]$ and $F$ is the Fisher matrix per trail,
\begin{equation}
F_{tr}=\frac{1}{N}\left\langle \frac{\partial \ln \mathcal{L}}{\partial
p_{0t}}\frac{\partial \ln \mathcal{L}}{\partial p_{0r}}\right\rangle ,
\label{fisher}
\end{equation}%
being $\mathcal{L}=\prod_{i=0}^{2}P_{i}(\mathbf{n_{i}|\mathbf{p_{i}}})$ the
likelihood. After straightforward but lengthy calculations (see Appendix) we
find that the lower bound for the estimation error per trial in terms of
measured probabilities is given by
\begin{eqnarray}
\left\langle \mathcal{E}_{min}^{2}\right\rangle  &=&\frac{2}{1-\lambda ^{2}}%
\left[ \left( (1+\lambda ^{2})p_{00}p_{10}+p_{01}p_{11}+p_{02}p_{12}\right)
\right.   \notag \\
&&\left. -\frac{4\lambda ^{2}}{S^{2}}p_{00}p_{10}\left(
p_{01}p_{11}+p_{02}p_{12}\right) \right] .  \label{error2}
\end{eqnarray}%
It is easy to see that at $\lambda =0$ (corresponding to $\theta =\pi /4$)
the mean Hilbert-Schmidt distance for MUBs is recovered \cite{checos}.

The lower bound (\ref{error2}) can still be averaged over the space of
quantum states. We will consider pure and mixed states  separately.

Let us first consider an arbitrary pure state $\left\vert \psi \right\rangle
$ with projections $x$ and $1-x$ on the basis $\{\left\vert \psi
_{k}^{0}\right\rangle ,k=0,1\}$, that can be taken as $x=|\langle \psi |\psi
_{0}^{0}\rangle |^{2}=p_{00}$ due to invariance of the averaging procedure
under unitary transformations. It is straightforward to check that
\begin{equation}
p_{01}p_{11}+p_{02}p_{12}=\frac{S^{2}}{2}-(1-\lambda ^{2})x(1-x).  \notag
\end{equation}%
Thus, the averaged, over the space of pure states, MSE takes the form
\begin{eqnarray}
\left\langle \langle \mathcal{E}_{min}^{2}\rangle \right\rangle
&=&\left\langle \frac{S^{2}}{1-\lambda ^{2}}+\frac{8\lambda
^{2}x^{2}(1-x)^{2}}{S^{2}}\right\rangle   \notag \\
&=&\frac{3+\lambda ^{2}}{3(1-\lambda ^{2})}+\frac{2(3-2\lambda ^{2})}{%
3\lambda ^{2}}  \notag \\
&&-\frac{2(1-\lambda ^{2})}{\lambda ^{3}}\arctanh(\lambda ),  \label{erroran}
\end{eqnarray}%
where the double brackets mean averaging both over the sample and over the
space of states. For $\lambda =0$, corresponding to the standard MUB
tomography, $\langle \langle \mathcal{E}_{min}^{2}\rangle \rangle =1$ \cite%
{checos}, while in the limit $\lambda \rightarrow 1$ the lower bound of the
MSE diverges as $\left( 1-\lambda ^{2}\right) ^{-1}$, which qualitatively
coincides with results of \cite{vallone}.

\begin{figure}[h]
\begin{center}
\includegraphics[width=8cm]{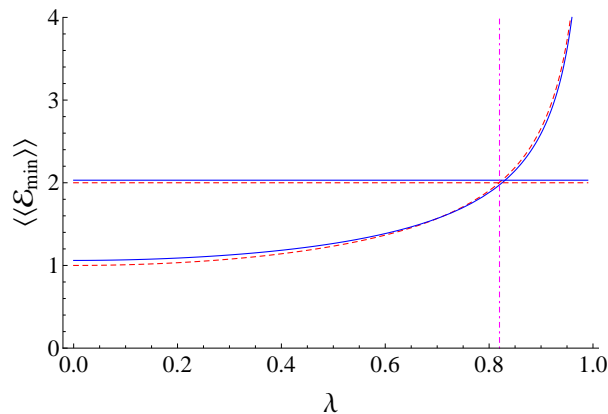}
\end{center}
\caption{(Colour online) The average $\langle \langle \mathcal{E}_{min}(%
\protect\lambda )\rangle \rangle $ over a sample of $10^{5}$ random statesas
a fucntion of $\protect\lambda =\cos 2\protect\theta $: (red) dashed line
for pure states, (blue) continuous line for mixed states. Average error for
SIC-POVM does not depend on $\protect\lambda $ and is represented as a
constant (red) dashed line $\langle\langle \mathcal{E}_{min}\rangle\rangle =2$ for pure
states and (blue) continuous line $\langle\langle \mathcal{E}_{min}\rangle\rangle =2.04$
for random mixed states. The vertical line at $\protect\lambda =0.82$ shows
the bound where the $\langle \langle \mathcal{E}_{min}(\protect\lambda %
)\rangle \rangle \leq \langle\langle \mathcal{E}_{min}\rangle\rangle _{SIC-POVM}$.}
\label{error1qb}
\end{figure}

In order to average over mixed states we use the eigenvalue distribution
based on the Bures metric \cite{bengtsson,hall},
\begin{equation}
p(x)=\frac{2}{\pi }\frac{(1-2x)^{2}}{\sqrt{x(1-x)}}.  \notag
\end{equation}%
Making use of the spectral decomposition $\rho =x\left\vert \rho
_{0}\right\rangle \left\langle \rho _{0}\right\vert +(1-x)\left\vert \rho
_{1}\right\rangle \left\langle \rho _{1}\right\vert $, where the eigenstates
can be parametrized as
\begin{eqnarray}
\left\vert \rho _{0}\right\rangle  &=&\cos \delta /2\left\vert \psi
_{0}^{0}\right\rangle +e^{i\eta _{0}}\sin \delta /2\left\vert \psi
_{1}^{0}\right\rangle   \notag \\
\left\vert \rho _{1}\right\rangle  &=&\sin \delta /2\left\vert \psi
_{0}^{0}\right\rangle +e^{i\eta _{1}}\cos \delta /2\left\vert \psi
_{1}^{0}\right\rangle ,  \notag
\end{eqnarray}%
with $\delta \in \lbrack 0,\pi ]$, and $\eta _{0},\eta _{1}\in \lbrack
0,2\pi ]$ we perform integration of (\ref{error2}) with the measure $%
p(x)\sin \delta dxd\delta d\eta _{0}d\eta _{1}/(8\pi ^{2})$. The result of
such integration can be found analytically in terms of special functions and
studied in the limit cases. Due to its cumbersome form we do not present
the explicit expression, but instead plot it in Fig. \ref{error1qb}.

In Fig.~\ref{error1qb} we plot $\langle \langle \mathcal{E}_{min}\rangle
\rangle =\sqrt{\langle \langle \mathcal{E}_{min}^{2}\rangle \rangle }$,
where the average is taken for a sample of $10^{5}$ pure and mixed random
states following the routine introduced in \cite{math}. For pure states, the
plot of the square root of equation (\ref{erroran}) perfectly coincides with
the numerical results. The mixed states are produced according to the Bures
metric. As it is expected, the best estimation is obtained for MUB
tomography, with $\langle \langle \mathcal{E}_{min}\rangle \rangle =1$ for
pure states, and $\langle \langle \mathcal{E}_{min}\rangle \rangle \approx
1.12$ for mixed states. One can also clearly see that the stronger the
measurements are,  the smaller  the estimation errors are.

The performance of DST can be also compared with a tomographic scheme based
on symmetric informationally complete positive operator valued measure
(SIC-POMV) measurements \cite{sic}. For a single qubit a set of projectors $\{\Pi _{k},k=1,..,4\}$
such that $\Tr(\Pi _{k}\Pi _{l})=1/3,k\neq l$ and $\sum_{k=1}^{4}\Pi _{k}=I,$
span the density matrix%
\begin{equation}
\rho _{s}=3\sum_{k=1}^{4}p_{k}\Pi _{k}-I,
\end{equation}%
where the probabilities $p_{k}=\Tr(\rho \Pi _{k})/2$ are the outcomes
associated with measurement of the operator $\Pi _{k}$,  $\sum_{k=1}^{4}p_{k}=1$. The corresponding MSE lower bound  has the form (\ref{mse}), where the
components of the matrix $Q$ are $q_{kl}=6\left( 1+\delta _{kl}\right) $, $%
k,l=1,2,3$, and the Fisher matrix elements per trail are $F=1/p_{4}+\delta
_{kl}/p_{k},$ which leads to $\langle \langle \mathcal{E}_{min}\rangle
\rangle =2$ for pure states \cite{englert}. In Fig.\ref{error1qb} we plot $%
\langle \langle \mathcal{E}_{min}\rangle \rangle $ for SIC-POVM tomography
as (red) dashed constant line\ for pure states and  as a (blue) continuous
constant line for mixed states, produced according to the Bures metric,
obtaining in this case $\langle \langle \mathcal{E}_{min}\rangle \rangle
=2.04$\  by averaging over $10^{5}$ randomly generated states. One can
observe that DST outperforms SIC-POVM qubit tomography for $\lambda <0.82$,
which is indicated in Fig.~\ref{error1qb} as a vertical (magenta)
dotted-dashed line.

\section{Conclusions}

We have shown that the performance of the DST protocol can be analyzed in a
similar way as in the standard projection-based reconstruction schemes. In
the framework of our approach we have been able to determine the estimation
error for any measurement strength, including the weak measurement case. In
addition, an explicit analytic form for the minimum square error have been
found in the pure and mixed states. The proposed scheme can be extended to
higher dimensions and composite many-particle systems.

\section{Appendix: MSE lower bound for any strength measurement}

In this Appendix we briefly deduce Eq. (\ref{error2}). Taking into account
the overlaps \cite{pra}
\begin{eqnarray}
|\langle \phi _{k}^{i}|\phi _{l}^{i}\rangle |^{2} &=&(1-\lambda ^{2})\delta
_{k,l}+\lambda ^{2},~~i=1,2  \notag \\
|\langle \phi _{k}^{i}|\phi _{l}^{j}\rangle |^{2} &=&\frac{1+\lambda ^{2}}{2}%
~~i\neq j,k,l=0,1  \notag \\
|\langle \phi _{0}^{0}|\phi _{l}^{i}\rangle |^{2} &=&\frac{1-\lambda }{2}%
~~i,j=1,2,k,l=0,1  \notag \\
|\langle \phi _{1}^{0}|\phi _{l}^{i}\rangle |^{2} &=&\frac{1+\lambda }{2}%
~~i,j=1,2,k,l=0,1,  \notag
\end{eqnarray}
and substituting the restrictions $\Delta p_{10}=-\Delta p_{00}$, $\Delta
p_{1i}=2\lambda \Delta p_{00}-\Delta p_{0i}$ into (\ref{E}) and (\ref%
{reconstruccion}) one arrives to
\begin{equation*}
Q=\frac{2}{1-\lambda ^{2}}\left(
\begin{array}{ccc}
1+\lambda ^{2} & -\lambda  & -\lambda  \\
-\lambda  & 1 & 0 \\
-\lambda  & 0 & 1%
\end{array}%
\right) .
\end{equation*}

The Fisher matrix (per trial) (\ref{fisher}) is obtained directly form the
likelihood
\begin{equation}
\mathcal{L}=\frac{1}{S^{2N}}\prod_{i=0}^{2}\frac{N!}{n_{0i}!n_{1i}!}%
p_{0i}^{n_{0i}}p_{1i}^{n_{1i}}.  \notag
\end{equation}%
In particular, one has
\begin{equation}
F_{00}=\frac{1}{p_{00}}+\frac{1}{p_{10}}+\frac{4\lambda ^{2}}{S}\left( \frac{%
1}{p_{11}}+\frac{1}{p_{12}}\right) -\frac{8\lambda ^{2}}{S^{2}},  \notag
\end{equation}%
where the two first terms correspond to the MUB tomography \cite{checos},
the third term appears due to dependence of the sum of probabilities in the
non-orthogonal bases on $p_{00}$, and the last term comes form the
normalization factor $S^{-2N}$. The main difference with the MUB case
consists in appearing elements in $F$ outside of the main diagonal, which is
a consequence of the relation (\ref{s}):
\begin{equation}
F_{0t}=F_{t0}=-\frac{2\lambda }{Sp_{1t}},~~t=1,2  \notag
\end{equation}%
For the non-orthogonal bases $t,r=1,2$, the elements are similar to the MUB
case, normalized by the factor $S$:
\begin{equation}
F_{tr}=\frac{1}{S}\left( \frac{1}{p_{0t}}+\frac{1}{p_{1t}}\right) \delta
_{tr}.  \notag
\end{equation}%
Substituting the explicit forms of $Q$ and $F$ into (\ref{mse}) one obtains (%
\ref{error2}).

\acknowledgments This work is supported by the Grant 254127 CONACyT, Mexico.

\end{document}